# On the Possible Trajectories of Spinning Particles.
## II. Particles in the Stationary Homogeneous Magnetic Field[1]

### A. N. Tarakanov

*Institute of Informational Technologies*
*Belarusian State University of Informatics and Radioelectronics*
*Kozlov str. 28, 220037, Minsk, Belarus*

E-mail: tarak-ph@mail.ru

The behavior of spinning particles in the stationary homogeneous magnetic field is considered and all types of trajectories for massive and massless particles are found. It is shown that spin of particles in a magnetic field is always arranged parallel or antiparallel to the field. Helicity plays a role of electric charge. The oscillation frequency of massless particle in a magnetic field increases.

PACS numbers: 14.60Cd, 41.90+e, 45.20.–d, 45.50.–j, 85.75.–d

Keywords: Classical mechanics, Equations of motion, Spin, Electron, Massless particles, Magnetic field

## 1. Introduction

Knowing the trajectories of charged particles with spin in an external field is applied in many areas of physics, geo- and astrophysics. This makes possible to calculate the various parameters of particles which account is required in problems of accelerator technology, on the retention of particles in magnetic traps arising in controlled thermonuclear fusion research, in plasma physics, magneto-optical studies, etc.

In the first part of this research [1] the equations of motion of spinning particles in any external field are obtained, and solutions for free particles are found under the assumption that potential function depends only on the velocity of particle relative to its center of inertia. The next challenge is to find solutions for particles that move in electric and magnetic fields. This paper is the second part dealing with motion of particles in a stationary homogeneous magnetic field. In Sec. 2 the equations of motion in an external field, which are somewhat different for massive and massless particles, are written in the Frenet-Serret basis, that allows to consider the motion of spinning particles both in stationary and non-stationary fields. Sec. 3 concerns the behavior of massive spinning particles in stationary homogeneous magnetic field, and Sec. 4 studies the same behavior of massless particles. We find all types of trajectories in both cases. It is shown that spin of particles in a magnetic field is always arranged parallel or antiparallel to the field, and the oscillation frequency of massless particle in a magnetic field increases.

## 2. Equations of motion of spinning particle in external field in moving reference frame

The form of equation of motion of spinning particle in an external field according to [1] depends on the relationship between the electric field strength and potential, which can be defined in two ways. If we assume the definition

$$\mathbf{E} = -\frac{\partial U}{\partial \mathbf{R}} + \frac{d}{dt}\left(\frac{\partial U}{\partial \mathbf{V}} - [\mathbf{S}^{\text{ext}} \times \dot{\mathbf{V}}]\right), \quad ([1], \text{Eq. } (2.9)), \qquad (2.1)$$

then the equation of motion takes the form

$$\frac{d}{dt}\left\{m_0 \mathbf{V} + \varsigma[\mathbf{s} \times \dot{\mathbf{V}}]\right\} = \mathbf{E} + [\mathbf{V} \times (\mathbf{B} + \varsigma \Omega_0^2 \mathbf{s})], \quad ([1], \text{Eq. } (2.11)), \qquad (2.2)$$

and if

$$\mathbf{E} = -\frac{\partial U}{\partial \mathbf{R}}, \quad ([1], \text{Eq. } (2.21)), \qquad (2.3)$$

---

[1] Parts I ([1]) and II are extended report to Proceedings of 9th International Conference Bolyai-Gauss-Lobachevsky: Non-Euclidean Geometry In Modern Physics (BGL-9), 27-30 Oct 2015, Minsk, Belarus.

then

$$\frac{d}{dt}\left(m_0\mathbf{V} - \frac{\partial U}{\partial \mathbf{V}} + \varsigma[\mathbf{s}\times\dot{\mathbf{V}}] + [\mathbf{S}^{\text{ext}}\times\dot{\mathbf{V}}]\right) = \mathbf{E} + [\mathbf{V}\times(\mathbf{B} + \varsigma\Omega_0^2\mathbf{s})]. \quad (2.4)$$

It is easy to see that (2.2) is the special case of (2.4) when $\partial U / \partial \mathbf{V} = 0$, $\mathbf{S}^{\text{ext}} = \mathbf{0}$. Solution of the equation (2.4) is defined both by the dependence of the fields $\mathbf{E}$, $\mathbf{B}$ and $\mathbf{S}^{\text{ext}}$ and the form of potential function $U$, which remains for free particle, when these fields vanish. In the moving r. f. K′ equation (2.4) splits into two equations

$$\frac{d}{dt}\left(m_0\mathbf{v} - \frac{\partial u}{\partial \mathbf{v}} + \varsigma[\mathbf{s}\times\dot{\mathbf{v}}] + [\mathbf{S}^{\text{ext}}\times\dot{\mathbf{v}}]\right) = \mathbf{E} + [\mathbf{v}\times(\mathbf{B} + \varsigma\Omega_0^2\mathbf{s})], \quad (2.5)$$

$$\frac{d}{dt}\left(m_0\mathbf{V}_{(K')} + \varsigma[\mathbf{s}\times\dot{\mathbf{V}}_{(K')}] + [\mathbf{S}^{\text{ext}}\times\dot{\mathbf{V}}_{(K')}]\right) = [\mathbf{V}_{(K')}\times(\mathbf{B} + \varsigma\Omega_0^2\mathbf{s})], \quad (2.6)$$

where $\mathbf{R} = \mathbf{R}_{(K')} + \mathbf{r}$, $\mathbf{V} = \mathbf{V}_{(K')} + \mathbf{v}$.

In the paper [1], where solutions for free particles are obtained, we assume that the potential function depends only on the velocity of particle relative to the center of inertia, $U = u(v)$. Here we shall assume that this condition is satisfied for stationary homogeneous fields. Then, decomposing all vectors and pseudo-vectors in (2.5) and (2.6) in Frenet-Serret basis ([1], Appendix) and choosing the binormal direction as the direction of the velocity of the r. f. K′, $\mathbf{V}_{(K')} = V_{(K')}\mathbf{e}_b$, we obtain a set of two vector equations

$$\left[\frac{d}{dt}\left(m_0 v - \frac{du}{dv} - (\varsigma s_b + S_b^{\text{ext}})v^2 K\right) - (\varsigma s_b + S_b^{\text{ext}})v\dot{v}K\right]\mathbf{e}_\tau +$$
$$+\left[\left(m_0 v - \frac{du}{dv}\right)vK + \dot{v}S_b^{\text{ext}} + (\varsigma s_b + S_b^{\text{ext}})(\ddot{v} - v^3 K^2)\right]\mathbf{e}_n -$$
$$-\left[(\varsigma s_\tau + S_\tau^{\text{ext}})v^2 K - (\varsigma s_n + S_n^{\text{ext}})\dot{v}\right]vT\mathbf{e}_n + \quad (2.7)$$
$$+\left[v^2 K\dot{S}_\tau^{\text{ext}} + (\varsigma s_\tau + S_\tau^{\text{ext}})(2v\dot{v}K + v^2\dot{K}) - \dot{v}\dot{S}_n^{\text{ext}} - (\varsigma s_n + S_n^{\text{ext}})\ddot{v} + (\varsigma s_b + S_b^{\text{ext}})v\dot{v}T\right]\mathbf{e}_b =$$
$$= E_\tau\mathbf{e}_\tau + (E_n - vB_b - \varsigma s_b\Omega_0^2 v)\mathbf{e}_n + (E_b + vB_n + \varsigma s_n\Omega_0^2 v)\mathbf{e}_b,$$

$$(\varsigma s_n + S_n^{\text{ext}})\ddot{V}_{(K')}\mathbf{e}_\tau + \left[(\varsigma s_\tau + S_\tau^{\text{ext}})vK + \dot{S}_n^{\text{ext}} + (\varsigma s_b + S_b^{\text{ext}})vT\right]\dot{V}_{(K')}\mathbf{e}_\tau +$$
$$+\left[\dot{S}_b^{\text{ext}}vT + (\varsigma s_b + S_b^{\text{ext}})(\dot{v}T + v\dot{T})\right]V_{(K')}\mathbf{e}_\tau - (\varsigma s_\tau + S_\tau^{\text{ext}})\ddot{V}_{(K')}\mathbf{e}_n +$$
$$+\left[(\varsigma s_n + S_n^{\text{ext}})vK - \dot{S}_\tau^{\text{ext}}\right]\dot{V}_{(K')}\mathbf{e}_n + \left[(\varsigma s_b + S_b^{\text{ext}})vK + (\varsigma s_\tau + S_\tau^{\text{ext}})vT - m_0\right]vTV_{(K')}\mathbf{e}_n + \quad (2.8)$$
$$+\left[m_0 - 2(\varsigma s_\tau + S_\tau^{\text{ext}})vT\right]\dot{V}_{(K')}\mathbf{e}_b - \left[\dot{S}_\tau^{\text{ext}}vT + (\varsigma s_\tau + S_\tau^{\text{ext}})(\dot{v}T + v\dot{T})\right]V_{(K')}\mathbf{e}_b =$$
$$= -(B_n + \varsigma s_n\Omega_0^2)V_{(K')}\mathbf{e}_\tau + (B_\tau + \varsigma s_\tau\Omega_0^2)V_{(K')}\mathbf{e}_n,$$

where $K$ and $T$ are curvature and torsion of trajectory, respectively.

The unit vectors of moving basis $\mathbf{e}_\tau$, $\mathbf{e}_n$ and $\mathbf{e}_b$ move in space and precess with an angular velocity represented by the Darboux vector $\mathbf{\Omega}^D = v(T\mathbf{e}_\tau + K\mathbf{e}_b)$. They are related with the unit vectors $\mathbf{e}_X$, $\mathbf{e}_Y$ and $\mathbf{e}_Z$ through rotation matrix, which may be parameterized in various ways. In many cases this relationship can be represented as

$$\mathbf{e}_\tau = \cos\Theta(t)\cos\Phi(t)\mathbf{e}_X + \cos\Theta(t)\sin\Phi(t)\mathbf{e}_Y + \sin\Theta(t)\mathbf{e}_Z. \quad (2.9)$$

$$\mathbf{e}_n = -\frac{\dot{\Theta}\sin\Theta\cos\Phi + \dot{\Phi}\cos\Theta\sin\Phi}{\sqrt{\dot{\Theta}^2 + \dot{\Phi}^2\cos^2\Theta}}\mathbf{e}_X - \frac{\dot{\Theta}\sin\Theta\sin\Phi - \dot{\Phi}\cos\Theta\cos\Phi}{\sqrt{\dot{\Theta}^2 + \dot{\Phi}^2\cos^2\Theta}}\mathbf{e}_Y +$$
$$+\frac{\dot{\Theta}\cos\Theta}{\sqrt{\dot{\Theta}^2 + \dot{\Phi}^2\cos^2\Theta}}\mathbf{e}_Z, \quad (2.10)$$

$$\mathbf{e}_{\mathrm{b}} = \frac{\dot{\Theta}\sin\Phi - \dot{\Phi}\sin\Theta\cos\Theta\cos\Phi}{\sqrt{\dot{\Theta}^2 + \dot{\Phi}^2\cos^2\Theta}}\mathbf{e}_X - \frac{\dot{\Theta}\cos\Phi + \dot{\Phi}\sin\Theta\cos\Theta\sin\Phi}{\sqrt{\dot{\Theta}^2 + \dot{\Phi}^2\cos^2\Theta}}\mathbf{e}_Y +$$
$$+ \frac{\dot{\Phi}\cos^2\Theta}{\sqrt{\dot{\Theta}^2 + \dot{\Phi}^2\cos^2\Theta}}\mathbf{e}_Z. \tag{2.11}$$

This choice corresponds to representation of the relative velocity as
$$\mathbf{v} = v(t)[\cos\Theta(t)\cos\Phi(t)\mathbf{e}_X + \cos\Theta(t)\sin\Phi(t)\mathbf{e}_Y + \sin\Theta(t)\mathbf{e}_Z]. \tag{2.12}$$

If we choose the direction of binormal fixed, such as in the case of free particles or particles in a homogeneous external field, then it follows from equation $\dot{\mathbf{e}}_{\mathrm{b}} = -vT\mathbf{e}_{\mathrm{n}} = 0$ that
$$T = \frac{(\mathbf{v}\cdot[\dot{\mathbf{v}}\times\ddot{\mathbf{v}}])}{[\mathbf{v}\times\dot{\mathbf{v}}]^2} = \frac{(\dot{\Phi}\ddot{\Theta} - \ddot{\Phi}\dot{\Theta})\cos\Theta + (2\dot{\Theta}^2 + \dot{\Phi}^2\cos^2\Theta)\dot{\Phi}\sin\Theta}{v(\dot{\Theta}^2 + \dot{\Phi}^2\cos^2\Theta)} = 0, \tag{2.13}$$

which shows that the torsion $T$ vanishes, if $\Theta = 0$ or $\dot{\Phi} = 0$, and corresponds to the representation of the relative velocity and the velocity of r. f. $K'$ in the form
$$\mathbf{v}(t) = v(t)\mathbf{e}_\tau = v(t)[\cos\Phi(t)\mathbf{e}_X + \sin\Phi(t)\mathbf{e}_Y], \quad \mathbf{V}_{(K')}(t) = V_{(K')}(t)\mathbf{e}_{\mathrm{b}} = V_{(K')}(t)\mathbf{e}_Z. \tag{2.14}$$

Then curvature is
$$K = \frac{|[\mathbf{v}\times\dot{\mathbf{v}}]|}{v^3} = \frac{\sqrt{\dot{\Theta}^2 + \dot{\Phi}^2\cos^2\Theta}}{v} = \frac{\dot{\Phi}}{v}. \tag{2.15}$$

Equation of motion of spin
$$\mathbf{s} = (s_\tau\cos\Phi - s_{\mathrm{n}}\sin\Phi)\mathbf{e}_X + (s_\tau\sin\Phi + s_{\mathrm{n}}\cos\Phi)\mathbf{e}_Y + s_{\mathrm{b}}\mathbf{e}_Z \tag{2.16}$$

becomes
$$\dot{\mathbf{s}} = [\mathbf{\Omega}_{\mathrm{D}}\times\mathbf{s}] = \dot{\Phi}\Big[-(s_\tau\sin\Phi + s_{\mathrm{n}}\cos\Phi)\mathbf{e}_X + (s_\tau\cos\Phi - s_{\mathrm{n}}\sin\Phi)\mathbf{e}_Y\Big], \tag{2.17}$$

where
$$\mathbf{\Omega}_{\mathrm{D}} = \dot{\Phi}\mathbf{e}_Z. \tag{2.18}$$

Equation of motion (2.4) at $U = u(v)$ for any external fields $\mathbf{E}$, $\mathbf{B}$ and $\mathbf{S}^{\mathrm{ext}}$ leads to conservation of total energy. In general, such a dependence may be insufficient, and its definition is a matter of a separate study. Equations of motion (2.7), (2.8) should be then modified in accordance with the form of potential function.

### 3. Massive spinning particle in stationary homogeneous magnetic field

Taking into account the above assumptions about potential function and the moving r. f. $K'$, equations of motion of spinning particle in magnetic field (at $\mathbf{E} = \mathbf{0}$, $\mathbf{S}^{\mathrm{ext}} = \mathbf{0}$) reduce to the set of equations
$$\left(m_0 v - \frac{du}{dv}\right)\dot{\Phi} + \varsigma s_{\mathrm{b}}[\ddot{v} + (\Omega_0^2 - \dot{\Phi}^2)v] = -B_{\mathrm{b}}v, \tag{3.1}$$
$$\varsigma s_\tau(v\ddot{\Phi} + \dot{v}\dot{\Phi}) - \varsigma s_{\mathrm{n}}(\ddot{v} + \Omega_0^2 v) = B_{\mathrm{n}}v, \tag{3.2}$$

and $\dot{V}_{(K')} = 0$,
$$B_{\mathrm{n}} = -B_X\sin\Phi + B_Y\cos\Phi = -\varsigma s_{\mathrm{n}}\Omega_0^2, \tag{3.3}$$
$$B_\tau = B_X\cos\Phi + B_Y\sin\Phi = -\varsigma s_\tau\Omega_0^2, \tag{3.4}$$

for massive particles or $\dot{V}_{(K')} \neq 0$,
$$\varsigma s_{\mathrm{n}}(\ddot{V}_{(K')} + \Omega_0^2 V_{(K')}) + \varsigma s_\tau\dot{\Phi}\dot{V}_{(K')} = -B_{\mathrm{n}}V_{(K')} = (B_X\sin\Phi - B_Y\cos\Phi)V_{(K')}, \tag{3.5}$$
$$\varsigma s_\tau(\ddot{V}_{(K')} + \Omega_0^2 V_{(K')}) - \varsigma s_{\mathrm{n}}\dot{\Phi}\dot{V}_{(K')} = -B_\tau V_{(K')} = -(B_X\cos\Phi + B_Y\sin\Phi)V_{(K')} \tag{3.6}$$

for massless particles.

Furthermore, it follows from the expression of the self-energy ([1], Eq. (2.49)) that

$$\varsigma s_b \dot{\Phi} = \frac{m_0}{2} - \frac{d}{dv}\frac{u}{v} - \frac{\mathcal{E}_0}{v^2}, \tag{3.7}$$

$$\varsigma s_b (v\ddot{\Phi} + 2\dot{v}\dot{\Phi}) = \frac{d}{dt}\left(m_0 v - \frac{du}{dv}\right). \tag{3.8}$$

Substituting (3.1) into (3.8) leads to the equation

$$\frac{d}{dt}\frac{[\ddot{v} + (\Omega_0^2 + B_b / \varsigma s_b)v]}{\dot{\Phi}} + \dot{v}\dot{\Phi} = 0, \tag{3.9}$$

which implies the first integral

$$\frac{[\ddot{v} + (\Omega_0^2 + B_b / \varsigma s_b)v]^2}{\dot{\Phi}^2} + \dot{v}^2 + (\Omega_0^2 + B_b / \varsigma s_b)v^2 = D^2, \tag{3.10}$$

where constant $D^2$ is always positive, if $B_b / \varsigma s_b \geq 0$, and may be non-positive, if $B_b / \varsigma s_b < 0$.

Substituting (3.3) into (3.2), we obtain

$$s_n \ddot{v} = s_\tau (v\ddot{\Phi} + \dot{v}\dot{\Phi}). \tag{3.11}$$

For homogeneous magnetic field its direction may be chosen as the direction of the velocity of moving r. f. Then $\mathbf{B} = B_b \mathbf{e}_b = B_Z \mathbf{e}_Z$, $B_Z > 0$, and (3.3)-(3.4) lead to $s_\tau = s_n = 0$, i. e. magnetic field $\mathbf{B}$ and spin $\mathbf{s}$ are collinear to the Z-axis, and equation (3.11) becomes identity. We put $s_b = es$, where $e$ is helicity, $e = +1$ for $s_b > 0$ and $e = -1$ for $s_b < 0$; $e = -1$ corresponds to anti-parallel direction of the field and spin (*electron state*), and $e = +1$ corresponds to parallel direction of the field and spin (*positron state*). Thus, helicity $e$ plays a role of electric charge.

Equations (3.7) and (3.9) are two equations for three unknown functions $\Phi(t)$, $v(t)$ and potential function $u(v)$, which should satisfy a separate equation. Since such equation is absent at present, we consider here a particular solution, corresponding to constant cyclotron frequency $\dot{\Phi} = \Omega_B = \mathrm{const}$. From (3.7) we find potential function

$$u(v) = \frac{(m_0 - 2\varsigma s_b \Omega_B)v^2}{2} + C_1 v + \mathcal{E}_0, \tag{3.12}$$

and the first integral of the equation (3.9) takes the form

$$\ddot{v} + (\Omega_0^2 + \Omega_B^2 + B_Z / \varsigma s_b)v = C_1 \Omega_B, \tag{3.13}$$

which yields several types of solutions defined by relationship between mass, and spin, and magnetic field.

**M.1.** $\Phi = \Omega_B t + \Phi_B$, $\Omega^2 = \Omega_0^2 + B_Z / \varsigma s_b > 0$. Equation (3.13) has the following solution

$$v(t) = v_1 + v_0 \sin(\Omega t + \varphi_0), \tag{3.14}$$

whence

$$\mathbf{v}(t) = [v_1 + v_0 \sin(\Omega t + \varphi_0)][\cos(\Omega_B t + \Phi_B)\mathbf{e}_X + \sin(\Omega_B t + \Phi_B)\mathbf{e}_Y]. \tag{3.15}$$

The relevant trajectory is described by radius vector

$$\begin{aligned}\mathbf{R}(t) = \mathbf{R}(0) &+ \left[\rho_B \cos\varphi_0 \cos\Phi_B - \rho(0)\sin\Phi_B\right]\mathbf{e}_X + \\ &+ \left[\rho_B \cos\varphi_0 \sin\Phi_B + \rho(0)\cos\Phi_B\right]\mathbf{e}_Y + \\ &+ \left[\rho(t)\sin(\Omega_B t + \Phi_B) - \rho_B \cos(\Omega t + \varphi_0)\cos(\Omega_B t + \Phi_B)\right]\mathbf{e}_X - \\ &- \left[\rho(t)\cos(\Omega_B t + \Phi_B) + \rho_B \cos(\Omega t + \varphi_0)\sin(\Omega_B t + \Phi_B)\right]\mathbf{e}_Y + V_{(K')}t\mathbf{e}_Z,\end{aligned} \tag{3.16}$$

where $\Omega = \Omega_B \sqrt{1 + \eta_B}$,

$$v_1 = \frac{C_1}{\Omega_B(1+\eta_B)}, \quad \eta_B = \frac{\Omega_0^2 + B_Z/\varsigma s_b}{\Omega_B^2}, \tag{3.17}$$

$$\rho_B = \frac{\varsigma s_b \Omega v_0}{\varsigma s_b \Omega_0^2 + B_Z} = \frac{v_0\sqrt{1+\eta_B}}{\Omega_B \eta_B}, \quad \rho(t) = \frac{\rho_B}{\sqrt{1+\eta_B}}\left[\frac{v_1\eta_B}{v_0} - \sin(\Omega t + \varphi_0)\right]. \tag{3.18}$$

Types of trajectories in a plane orthogonal to the direction of motion of the center of inertia are presented in Fig. 1-3 (at $|\eta_B| \ll 1$), Fig. 4-6 (at $0 < 1+\eta_B < 1$) and Fig. 7-9 (at $\eta_B \gg 1$) for values $v_1/v_0 = 0$, $0 < v_1/v_0 < 1$ and $v_1/v_0 \gg 1$. At $\dot{\Phi} = \Omega_B = 0$ the particle oscillates in the plane (XY), and its center of inertia moves uniformly along the Z-axis.

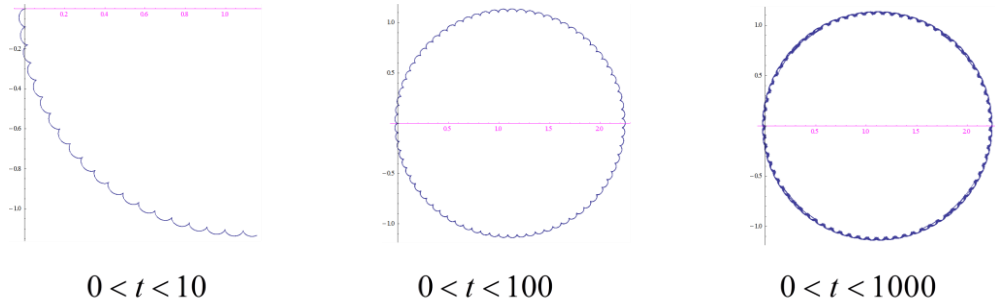

$0 < t < 10$   $0 < t < 100$   $0 < t < 1000$

Figure 1. Type of trajectories (3.16) of massive particle in magnetic field at $|\eta_B| \ll 1$ ($v_1/v_0 = 0$, $\Omega_B = 2,3$, $\eta_B = -0,05$)

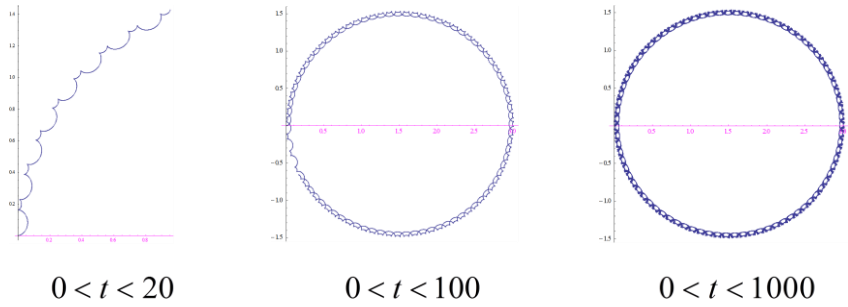

$0 < t < 20$   $0 < t < 100$   $0 < t < 1000$

Figure 2. Type of trajectories (3.16) of massive particle in magnetic field at $|\eta_B| \ll 1$ ($v_1/v_0 = 0,6$, $\Omega_B = 2,3$, $\eta_B = -0,05$)

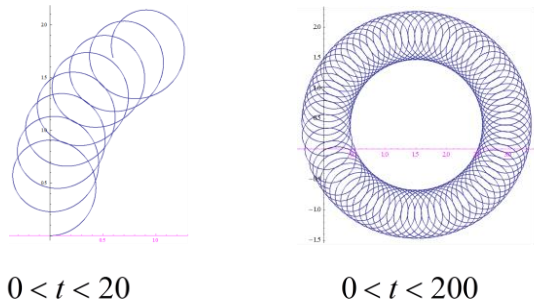

$0 < t < 20$   $0 < t < 200$

Figure 3. Type of trajectories (3.16) of massive particle in magnetic field at $|\eta_B| \ll 1$ ($v_1/v_0 = 10,0$, $\Omega_B = 2,3$, $\eta_B = -0,05$)

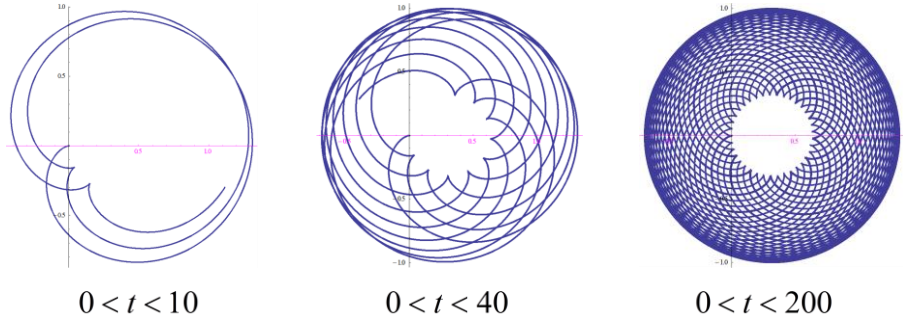

| $0 < t < 10$ | $0 < t < 40$ | $0 < t < 200$ |

Figure 4. Type of trajectories (3.16) of massive particle in magnetic field
at $0 < 1 + \eta_B < 1$ ($v_1/v_0 = 0$, $\Omega_B = 2,3$, $\eta_B = -0,9$)

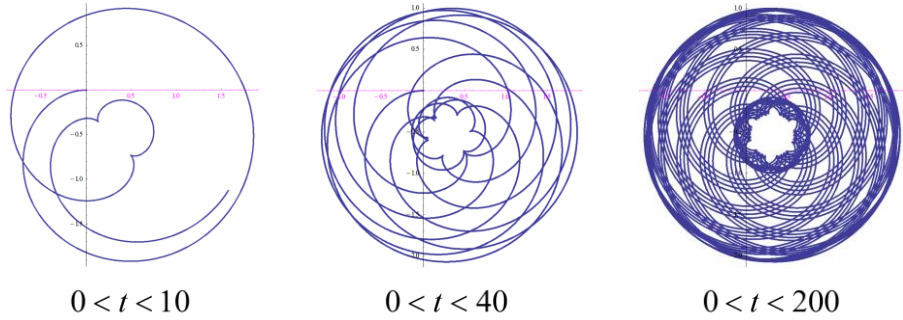

| $0 < t < 10$ | $0 < t < 40$ | $0 < t < 200$ |

Figure 5. Type of trajectories (3.16) of massive particle in magnetic field
at $0 < 1 + \eta_B < 1$ ($v_1/v_0 = 0,6$, $\Omega_B = 2,3$, $\eta_B = -0,9$)

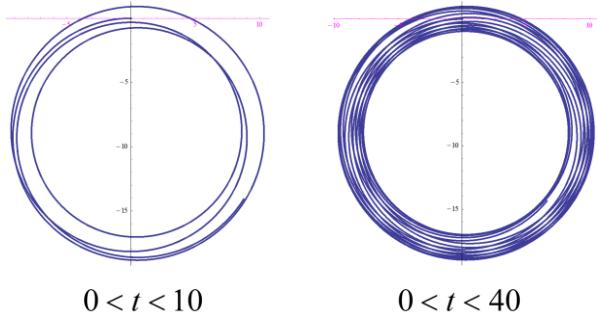

| $0 < t < 10$ | $0 < t < 40$ |

Figure 6. Type of trajectories (3.16) of massive particle in magnetic field
at $0 < 1 + \eta_B < 1$ ($v_1/v_0 = 10,0$, $\Omega_B = 2,3$, $\eta_B = -0,9$)

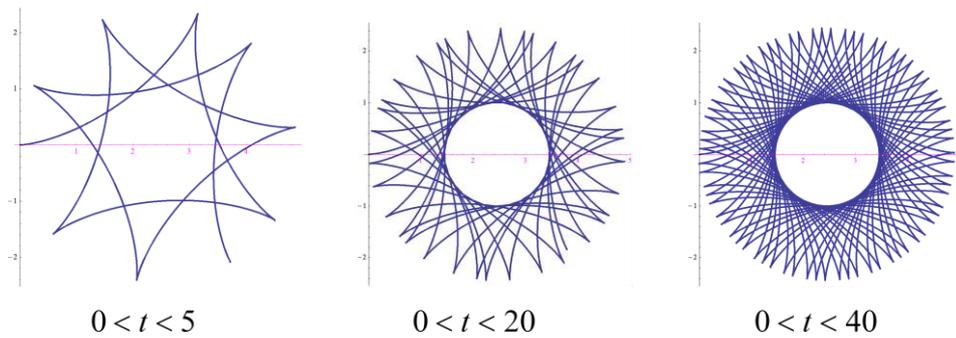

| $0 < t < 5$ | $0 < t < 20$ | $0 < t < 40$ |

Figure 7. Type of trajectories (3.16) of massive particle in magnetic field
at $\eta_B \gg 1$ ($v_1/v_0 = 0$, $\Omega_B = 2,3$, $\eta_B = 5,0$)

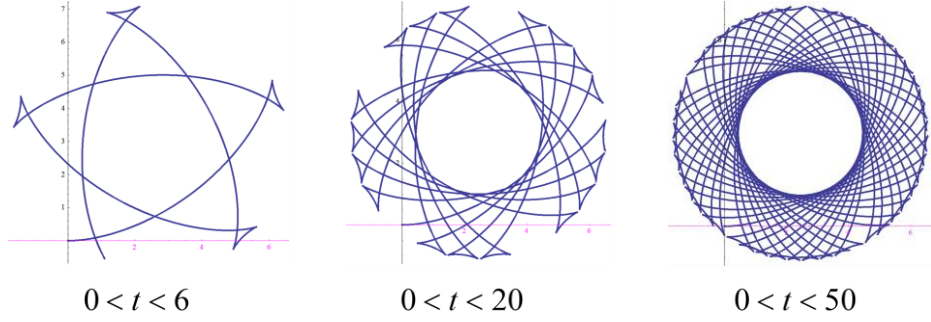

| $0 < t < 6$ | $0 < t < 20$ | $0 < t < 50$ |

Figure 8. Type of trajectories (3.16) of massive particle in magnetic field at $\eta_B \gg 1$ ($v_1/v_0 = 0,6$, $\Omega_B = 2,3$, $\eta_B = 5,0$)

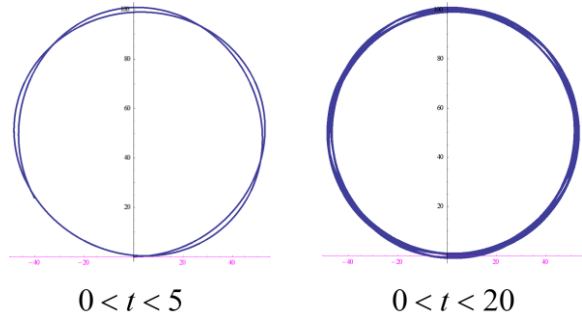

| $0 < t < 5$ | $0 < t < 20$ |

Figure 9. Type of trajectories (3.16) of massive particle in magnetic field at $\eta_B \gg 1$ ($v_1/v_0 = 10,0$, $\Omega_B = 2,3$, $\eta_B = 5,0$)

**M.2.** If the magnetic field is such that the relation

$$\Omega_0^2 = -\frac{B_Z}{\varsigma s_\mathrm{b}} \qquad (3.19)$$

is valid, i. e. $\eta_B = 0$, then $\Omega = \Omega_B$, $v_1 = C_1/\Omega_B$. Trajectory is presented by radius vector

$$\begin{aligned}\mathbf{R}(t) = \mathbf{R}(0) &+ \rho_0\left[\cos(2\Phi_B + \varphi_0) - k\sin\Phi_B\right]\mathbf{e}_X + \rho_0\left[\sin(2\Phi_B + \varphi_0) + k\cos\Phi_B\right]\mathbf{e}_Y - \\ &- \rho_0\left[\cos(2\Omega_B t + 2\Phi_B + \varphi_0) - k\sin(\Omega_B t + \Phi_B) - 2\Omega_B t\sin\varphi_0\right]\mathbf{e}_X - \\ &- \rho_0\left[\sin(2\Omega_B t + 2\Phi_B + \varphi_0) + k\cos(\Omega_B t + \Phi_B) - 2\Omega_B t\cos\varphi_0\right]\mathbf{e}_Y + V_{(\mathrm{K}')}t\mathbf{e}_Z,\end{aligned} \qquad (3.20)$$

where $\rho_0 = v_0/4\Omega_B$, $k = 4v_1/v_0$, and represents complicated curve in center-of-inertia r. f., which moves with velocity $\mathbf{V}_\mathrm{C} = (v_0/2)(\sin\varphi_0\mathbf{e}_X + \cos\varphi_0\mathbf{e}_Y) + V_{(\mathrm{K}')}\mathbf{e}_Z$. Examples of such paths are given in Fig. 10. At $\dot\Phi = \Omega_B = 0$ the law of motion (3.20) corresponds to uniform rectilinear

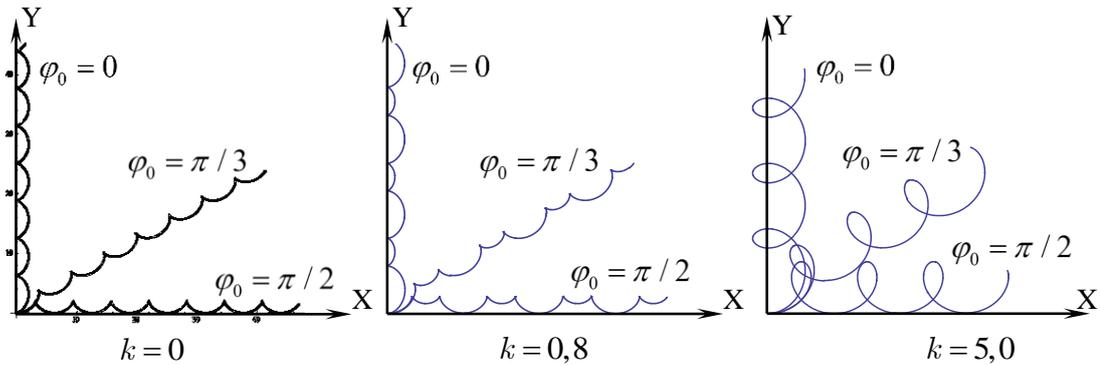

| $k = 0$ | $k = 0,8$ | $k = 5,0$ |

Figure 10. Type of trajectories (3.20) of massive particle in magnetic field at $\eta_B = 0$ ($\Omega_B = 2,3$, $\rho_0 = 1$, $0 < t < 10$)

movement along the Z-azis, $\mathbf{R}(t) = \mathbf{R}(0) + V_{(K')}t\mathbf{e}_Z$.

**M.3.** $\Phi = \Omega_B t + \Phi_B$, $\tilde{\Omega}^2 = -\Omega_0^2 - \Omega_B^2 - B_Z/\varsigma s_b > 0$, т. е. $\eta_B < -1$. In this case equation (3.13) gives

$$\mathbf{v}(t) = [v_1 + v_0 \,\text{sh}(\tilde{\Omega}t + \varphi_0)][\cos(\Omega_B t + \Phi_B)\mathbf{e}_X + \sin(\Omega_B t + \Phi_B)\mathbf{e}_Y], \quad (3.21)$$

$$\mathbf{R}(t) = \mathbf{R}(0) + \left[\tilde{\rho}_B \,\text{ch}\,\varphi_0 \cos\Phi_B - \tilde{\rho}(0)\sin\Phi_B\right]\mathbf{e}_X + \left[\tilde{\rho}_B \,\text{ch}\,\varphi_0 \sin\Phi_B + \tilde{\rho}(0)\cos\Phi_B\right]\mathbf{e}_Y +$$
$$+ \left[\tilde{\rho}(t)\sin(\Omega_B t + \Phi_B) - \tilde{\rho}_B \,\text{ch}(\tilde{\Omega}t + \varphi_0)\cos(\Omega_B t + \Phi_B)\right]\mathbf{e}_X - \quad (3.22)$$
$$- \left[\tilde{\rho}(t)\cos(\Omega_B t + \Phi_B) + \tilde{\rho}_B \,\text{ch}(\tilde{\Omega}t + \varphi_0)\sin(\Omega_B t + \Phi_B)\right]\mathbf{e}_Y + V_{(K')}t\mathbf{e}_Z,$$

where $\tilde{\Omega} = \Omega_B\sqrt{-1-\eta_B}$,

$$\tilde{\rho}_B = \frac{\varsigma s_b \tilde{\Omega} v_0}{\varsigma s_b \Omega_0^2 + B_Z} = \frac{v_0\sqrt{-1-\eta_B}}{\Omega_B \eta_B}, \quad \tilde{\rho}(t) = \frac{\tilde{\rho}_B}{\sqrt{-1-\eta_B}}\left[\frac{v_1 \eta_B}{v_0} - \text{sh}(\tilde{\Omega}t + \varphi_0)\right], \quad (3.23)$$

$\eta_B$ and $v_1$ are given in (3.17).

Trajectory (3.22) in the center-of-inertia r. f. is twisting ($t<0$), and then untwisting ($t>0$) helix (Fig. 11), and transforms into straight line at $\dot{\Phi} = \Omega_B = 0$.

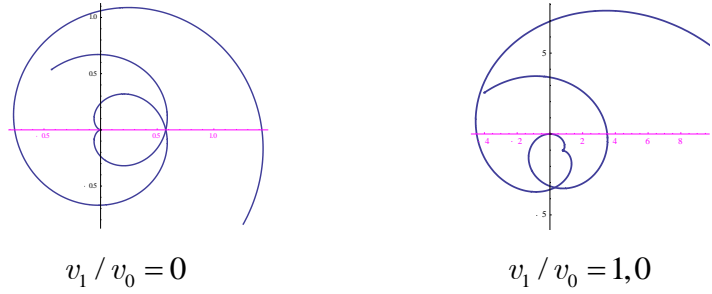

$v_1/v_0 = 0$          $v_1/v_0 = 1,0$

Figure 11. Type of trajectories (3.22) of massive particle in magnetic field
at $\eta_B < -1$ ($\eta_B = -1,1$, $\Omega_B = 2,3$, $-5 < t < 3$)

**M.4.** $\Omega_0^2 + \Omega_B^2 + B_Z/\varsigma s_b = 0$, i. e. cyclotron frequency is equal to

$$\Omega_B = \sqrt{-\Omega_0^2 - B_Z/\varsigma s_b}. \quad (3.24)$$

In this case (3.13) gives $v(t) = v_0 + wt + C_1\Omega_B t^2/2$,

$$\mathbf{R}(t) = \mathbf{R}(0) - \frac{1}{\Omega_B^2}\left[(v_0\Omega_B - C_1)\sin\Phi_B + w\cos\Phi_B\right]\mathbf{e}_X +$$
$$+ \frac{1}{\Omega_B^2}\left[(v_0\Omega_B - C_1)\cos\Phi_B - w\sin\Phi_B\right]\mathbf{e}_Y +$$
$$+ \frac{1}{\Omega_B^2}\left[(v(t)\Omega_B - C_1)\sin(\Omega_B t + \Phi_B) + \frac{dv(t)}{dt}\cos(\Omega_B t + \Phi_B)\right]\mathbf{e}_X + \quad (3.25)$$
$$+ \frac{1}{\Omega_B^2}\left[-(v(t)\Omega_B - C_1)\cos(\Omega_B t + \Phi_B) + \frac{dv(t)}{dt}\sin(\Omega_B t + \Phi_B)\right]\mathbf{e}_Y + V_{(K')}t\mathbf{e}_Z$$

The trajectories of this type at $w = 0$, $C_1 = 0$ correspond to the trajectories of classical Lorentz electrodynamics, where, as it is known, a charged particle, that is flying in uniform magnetic field, moves in a spiral or circle, when its velocity is perpendicular to field, and spin of the particle is not taken into account in no way. As it follows from the solutions obtained above, the spin of a particle that has fallen into magnetic field, has always arranged parallel or antiparal-lel to the field. This was conclusively proven by experiment of Stern and Gerlach. It follows from (3.24), that condition

$$B_Z < -\varsigma s_b \Omega_0^2 \tag{3.26}$$

should be satisfied. Assuming for the electron $\Omega_0 = m_e c^2 / \hbar \approx 7{,}77 \cdot 10^{20}$ Hz ([2], Eq. (4.50), or [3], Eq. (89)), $\varsigma = -c^{-2}$, $s_b = es = -\hbar/2$, we find limit value of magnetic field

$$B_{max} = m_e^2 c^2 / 2\hbar \approx 5{,}6 \cdot 10^{-11} \text{ kg/s}, \tag{3.27}$$

that corresponds to $B_{max} = m_e^2 c^2 / 2e\hbar \approx 3{,}5 \cdot 10^8$ T in SI. Large values of magnetic field are occurred in magnetars, that are neutron stars with strong magnetic field (up to $10^{11}$ T), wherein condition (3.26) is not valid. For such fields apparently can be realized cases M.1-M.3. For positron states ($e = +1$) condition (3.26) will be fulfilled, only if $\varsigma = +c^{-2}$. Therefore in general one can put $\varsigma = ec^{-2}$.

Finally, we get the standard solution assuming $s_b = 0$. Then (3.1) looks like

$$\left(m_0 v - \frac{du}{dv}\right)\dot{\Phi} = -vB_Z, \tag{3.28}$$

whence (at $\dot{\Phi} = \Omega_B$)

$$\frac{du}{dv} = (m_0 + B_Z/\Omega_B)v, \quad u(v) = \frac{1}{2}(m_0 + B_Z/\Omega_B)v^2. \tag{3.29}$$

$u(v) = 0$ corresponds to standard dependence of potential function from relative distance. As it follows from (3.29) this is possible, when

$$\Omega_B = -B_Z / m_0, \tag{3.30}$$

which coincides with standard definition of cyclotron frequency. Then the condition (3.26) is equivalent to $\Omega_B > \Omega_0 / 2$. For constant field (3.28), (3.29) give $v = v_0 = \text{const}$. Then trajectory is described by (3.25), or

$$\mathbf{R}(t) = \mathbf{R}(0) + \frac{v_0}{\Omega_B}\left[\sin(\Omega_B t + \Phi_B) - \sin\Phi_B\right]\mathbf{e}_X - \\ -\frac{v_0}{\Omega_B}\left[\cos(\Omega_B t + \Phi_B) - \cos\Phi_B\right]\mathbf{e}_Y + V_{(K')}t\mathbf{e}_Z. \tag{3.31}$$

Analyzing the solutions (3.16) (Fig. 1, 2, 4-6, 9) and (3.22) (Fig. 11), one can see that they are close to the classical solution (3.31).

**M.5.** $\dot{\Phi} \neq \text{const}$. Taking account of spin at $u(v) = 0$ reduces equations (3.7) and (3.9) to

$$\varsigma s_b \dot{\Phi} = \frac{m_0}{2} - \frac{\mathcal{E}_0}{v^2}, \tag{3.32}$$

$$\frac{d}{dt}\left[\frac{[\ddot{v} + (\Omega_0^2 + B_b / \varsigma s_b)v]v^2}{\frac{m_0 v^2}{2} - \mathcal{E}_0} + \frac{1}{\varsigma^2 s_b^2 v}\left(\frac{m_0 v^2}{2} + \mathcal{E}_0\right)\right] = 0. \tag{3.33}$$

From (3.33) we have

$$\ddot{v} + f(v) = 0, \tag{3.34}$$

where

$$f(v) = \Omega^2 v + \frac{C_2 \mathcal{E}_0}{v^2} - \frac{\mathcal{E}_0^2}{\varsigma^2 s_b^2 v^3} - \frac{C_2 m_0}{2}, \quad C_2 = \text{const}, \tag{3.35}$$

$$\Omega^2 = \Omega_0^2 + \frac{B_b}{\varsigma s_b} + \frac{m_0^2}{4\varsigma^2 s_b^2}. \tag{3.36}$$

Obviously, classical solution (3.31) corresponds to $\mathcal{E}_0 = 0$, $C_2 = 0$, $m_0 = 2\varsigma s_b \Omega_B$, i. e. $\Omega_B = m_0 c^2 / \hbar = \Omega_0$.

Equation (3.34) admits the first integral

$$\dot{v}^2 + 2\int f(v)dv = C_3 = \text{const}, \tag{3.37}$$

whence

$$\int_{v_0}^{v} \frac{vdv}{\sqrt{-\Omega^2 v^4 + C_2 m_0 v^3 + C_3 v^2 + 2C_2 \mathcal{E}_0 v - \mathcal{E}_0^2 / \varsigma^2 s_b^2}} = \pm t. \tag{3.38}$$

The result of integration is determined by the signs of constants $\Omega^2$, $C_2$, $C_3$, $\mathcal{E}_0$ and the relationship between them, as well as the roots of the equation

$$\begin{aligned}-\Omega^2 v^4 + C_2 m_0 v^3 + C_3 v^2 + 2C_2 \mathcal{E}_0 v - \mathcal{E}_0^2 / \varsigma^2 s_b^2 = \\ = -\Omega^2 (v - v_1)(v - v_2)(v - v_3)(v - v_4) = 0.\end{aligned} \tag{3.39}$$

The enumeration of all possible solutions of equation (3.34) is not possible here because of their great number and bulkiness.

### 4. Massless spinning particle in stationary homogeneous magnetic field

For massless particles we have a set of equations (3.1), (3.11) and (3.5)-(3.6), which in a stationary homogeneous magnetic field $\mathbf{B} = B_b \mathbf{e}_b = B_Z \mathbf{e}_Z$ reduce to

$$s_n(\ddot{V}_{(K')} + \Omega_0^2 V_{(K')}) + s_\tau \dot{\Phi} \dot{V}_{(K')} = 0, \tag{4.1}$$

$$s_\tau(\ddot{V}_{(K')} + \Omega_0^2 V_{(K')}) - s_n \dot{\Phi} \dot{V}_{(K')} = 0. \tag{4.2}$$

Equation (3.1) admits the first integral (3.10), whence

$$\dot{\Phi} = \frac{\ddot{v} + (\Omega_0^2 + B_Z / \varsigma s_b)v}{\sqrt{D^2 - \dot{v}^2 - (\Omega_0^2 + B_Z / \varsigma s_b)v^2}} = -\frac{1}{\dot{v}} \frac{d}{dt} \sqrt{D^2 - \dot{v}^2 - (\Omega_0^2 + B_Z / \varsigma s_b)v^2}. \tag{4.3}$$

Substituting (4.3) into (3.1) at $m_0 = 0$ we get an equation

$$\begin{aligned}\frac{\ddot{v} + (\Omega_0^2 + B_Z / \varsigma s_b)v}{\sqrt{D^2 - \dot{v}^2 - (\Omega_0^2 + B_Z / \varsigma s_b)v^2}} \frac{du}{dv} = \\ = \varsigma s_b \left[1 - \frac{[\ddot{v} + (\Omega_0^2 + B_Z / \varsigma s_b)v]v}{D^2 - \dot{v}^2 - (\Omega_0^2 + B_Z / \varsigma s_b)v^2}\right][\ddot{v} + (\Omega_0^2 + B_Z / \varsigma s_b)v],\end{aligned} \tag{4.4}$$

which is valid either at 1) $\ddot{v} + (\Omega_0^2 + B_Z / \varsigma s_b)v = 0$ leading to $\dot{\Phi} = 0$, or at 2) $\ddot{v} + (\Omega_0^2 + B_Z / \varsigma s_b)v \ne 0$, $\dot{\Phi} \ne 0$. As a result we have following solutions.

**M$_0$.1.** $m_0 = 0$, $\Phi = \Phi_B = \text{const}$, $\ddot{v} + (\Omega_0^2 + B_Z / \varsigma s_b)v = 0$, which gives

$$v(t) = \begin{cases} v_0 \cos(\Omega t + \varphi_0), & \Omega^2 = \Omega_0^2 + B_Z / \varsigma s_b > 0, \\ v_0, & \Omega_0^2 + B_Z / \varsigma s_b = 0, \\ v_0 \operatorname{ch}(\tilde{\Omega} t + \tilde{\varphi}_0), & \tilde{\Omega}^2 = -\Omega_0^2 - B_Z / \varsigma s_b > 0. \end{cases} \tag{4.5}$$

Equations (3.11) and (4.1) reduce to identity, if $s_n = 0$, or $\ddot{v} = 0$, $s_n \ne 0$, and (3.7)-(3.8) lead to $du/dv = \text{const}$. (4.1)-(4.2) give solution

$$V_{(K')}(t) = V_{(K')0} \cos(\Omega_0 t + \delta_0). \tag{4.6}$$

The law of motion is easily obtained by integrating the velocity vector

$$\mathbf{R}(t) = \mathbf{R}(0) + \int_0^t [\mathbf{v}(t) + V_{(K')}(t)\mathbf{e}_Z]dt. \quad (4.7)$$

An analysis of possible relations between the quantities in (4.7), as well as determination the conditions of closed trajectories is not difficult.

**M$_0$.2.** $m_0 = 0$, $\dot{\Phi} \neq 0$, $\ddot{v} + (\Omega_0^2 + B_Z/\varsigma s_b)v \neq 0$. Equations (4.1)-(4.2) lead to $s_\tau^2 + s_n^2 = 0$, that possible only when $s_\tau = s_n = 0$. Then (4.1)-(4.2) reduce to identities. Equation (3.1) has an infinite set of solutions, one of which corresponds to $\Phi = \Omega_D t + \Phi_D$, where $\Omega_D = \pm\sqrt{\Omega_D^2} = \text{const}$. In this case (3.10) admits the first integral $\Omega_D v + \sqrt{D^2 - \dot{v}^2 - (\Omega_0^2 + B_Z/\varsigma s_b)v^2} = F$, from which we find the equation for the velocity

$$\dot{v} = \pm\sqrt{D^2 - F^2 + 2F\Omega_D v - (\Omega_D^2 + \Omega_0^2 + B_Z/\varsigma s_b)v^2}. \quad (4.8)$$

Substituting (4.8) into (4.3) leads to the equation

$$\ddot{v} + (\Omega_D^2 + \Omega_0^2 + B_Z/\varsigma s_b)v = F\Omega_D, \quad (4.9)$$

whose solution is a function

$$v(t) = \frac{F\Omega_D}{\chi_B^2} + v_0\cos(\chi_B t + \varphi_0), \quad \chi_B = \sqrt{\Omega_D^2 + \Omega_0^2 + B_Z/\varsigma s_b}, \quad (4.10)$$

$$\mathbf{v}(t) = \left(\frac{F\Omega_D}{\chi_B^2} + v_0\cos(\chi_B t + \varphi_0)\right)\left[\cos(\Omega_D t + \Phi_D)\mathbf{e}_X + \sin(\Omega_D t + \Phi_D)\mathbf{e}_Y\right]. \quad (4.11)$$

It follows from (4.8) and (4.10) that the velocity may vary in the limits $v_{min} \leq v \leq v_{max}$, where

$$v_{min} = \frac{F\Omega_D}{\chi_B^2} - v_0 \geq 0, \quad v_{max} = \frac{F\Omega_D}{\chi_B^2} + v_0, \quad v_0 = \frac{\sqrt{D^2\chi_B^2 - F^2\Omega_0^2 - F^2 B_Z/\varsigma s_b}}{\chi_B^2}. \quad (4.12)$$

The law of motion is similar to the equation (3.22) from [1]:

$$\begin{aligned}\mathbf{R}(t) = \mathbf{R}(0) &+ \frac{F}{\chi_B^2}[\sin(\Omega_D t + \Phi_D) - \sin\Phi_D]\mathbf{e}_X + \frac{F}{\chi_B^2}[\cos\Phi_D - \cos(\Omega_D t + \Phi_D)]\mathbf{e}_Y + \\
&+ \frac{v_0(\chi_B - \Omega_D)}{2(\Omega_0^2 + B_Z/\varsigma s_b)}\Big[\sin[(\chi_B + \Omega_D)t + \Phi_D + \varphi_0] - \sin(\Phi_D + \varphi_0)\Big]\mathbf{e}_X + \\
&+ \frac{v_0(\chi_B + \Omega_D)}{2(\Omega_0^2 + B_Z/\varsigma s_b)}\Big[\sin[(\chi_B - \Omega_D)t - \Phi_D + \varphi_0] + \sin(\Phi_D - \varphi_0)\Big]\mathbf{e}_X - \\
&- \frac{v_0(\chi_B - \Omega_D)}{2(\Omega_0^2 + B_Z/\varsigma s_b)}\Big[\cos[(\chi_B + \Omega_D)t + \Phi_D + \varphi_0] - \cos(\Phi_D + \varphi_0)\Big]\mathbf{e}_Y + \\
&+ \frac{v_0(\chi_B + \Omega_D)}{2(\Omega_0^2 + B_Z/\varsigma s_b)}\Big[\cos[(\chi_B - \Omega_D)t - \Phi_D + \varphi_0] - \cos(\Phi_D - \varphi_0)\Big]\mathbf{e}_Y + \\
&+ \int_0^t V_{(K')}(t)dt\,\mathbf{e}_Z,\end{aligned} \quad (4.13)$$

where $V_{(K')}(t)$ has an arbitrary dependence on time.

Condition of closed trajectory in the r. f. K$'$, $\mathbf{v}(t+T) = \mathbf{v}(t)$, leads to the relation $\Omega_D T = 2m\pi$, whence it follows $m\chi_B = l\Omega_D$, or

$$m^2(\Omega_0^2 + B_Z/\varsigma s_b) = (l^2 - m^2)\Omega_D^2, \quad (4.14)$$

where $l = 1, 2, ...$, $m = -l+1, -l+2, ..., 0, 1, 2, ..., l-1$, and $m = 0$ corresponds to $\Omega_D = 0$, i. e. to oscillations along the X-axis with frequency $\chi_{B0} = \sqrt{\Omega_0^2 + B_Z/\varsigma s_b}$. Comparison of solutions (4.5) and (4.11) with the appropriate solutions for free massless particles (Eqs. (3.9) and (3.22) from [1]) shows that the magnetic field induces an increase of the oscillation frequency.